%% file: skeleton.tex
\documentclass[a4paper,11pt]{article}
\usepackage{pos}

\DeclareFontFamily{U}{mathx}{}
\DeclareFontShape{U}{mathx}{m}{n}{<-> mathx10}{}
\DeclareSymbolFont{mathx}{U}{mathx}{m}{n}
\DeclareMathAccent{\widehat}{0}{mathx}{"70}
\DeclareMathAccent{\widecheck}{0}{mathx}{"71}
\usepackage[export]{adjustbox}
\usepackage{tikz}
\usetikzlibrary{decorations.pathmorphing,decorations.pathreplacing,calligraphy,calc,snakes,cd,external,decorations.markings}
\usepackage{dirtytalk}
\usepackage{orcidlink}
\usepackage{bbold}
\usepackage{mathtools}
\usepackage{fmtcount}
\usepackage{tcolorbox}

\newcounter{wordref}
\newcommand{\labelword}[2]{%
    \edef\@currentlabel{#2}%
    \refstepcounter{wordref}%
    \label{#1}%
    #2%
}

\title{Looping the loops: a tale of elliptic dual Feynman integrals}

\author*[a,\orcidlink{0000-0002-7005-9652}]{Mathieu Giroux}
\author[b,\orcidlink{0000-0003-1186-4624}]{Andrzej Pokraka}
\author[c,\orcidlink{0000-0002-3328-499X}]{Franziska Porkert}
\author[d]{Yoann Sohnle}

\affiliation[a]{Department of Physics, 
McGill University, 
3600 Rue University, 
Montréal, 
QC Canada H3A 2T8}

\affiliation[b]{
	Department of Physics, 	
	Brown University, 	
	Providence, 	
	RI 02912, 
	USA}

\affiliation[c]{Bethe Center for Theoretical Physics, 
Universität Bonn, 
D-53115, 
Germany
}

\affiliation[d]{Upsala Department of Physics and Astronomy,
Uppsala University,
75108 Uppsala,
Sweden}

\emailAdd{mathieu.giroux2@mail.mcgill.ca}
\emailAdd{andrzej\_pokraka@brown.edu}
\emailAdd{fporkert@uni-bonn.de}
\emailAdd{yoann.sohnle@physics.uu.se}

\abstract{In this talk, we review a loop-by-loop approach used to generate differential equations for multi-scale (dual) Feynman integrals. We illustrate the method on a well-established example: the unequal mass elliptic sunrise.}

\FullConference{
RADCOR2023\\
28th May - 2nd June, 2023\\
Crieff, Scotland, UK\\}

\vspace{.2cm}

\begin{document}
\maketitle

\section{Introduction}\label{sec:intro}

It is a long-standing fact, reminiscent of the golden age of string theory, that (hyper-)elliptic curves and Calabi-Yau manifolds play a central role in various corners of mathematics and physics. Over the past few decades, a heroic effort has been made by various authors to demonstrate this in the context of particle physics (see \cite{Bourjaily:2022bwx,Morales:2022csr,Pogel:2022vat,Duhr:2022dxb,Duhr:2022pch,Cao:2023tpx,Marzucca:2023gto,Bern:2023ccb} and references therein.) In this context, Feynman diagrams are used to compute scattering amplitudes, 
which yields observables such as cross-sections measurable in collider experiments or waveforms detected in gravitational wave observatories.

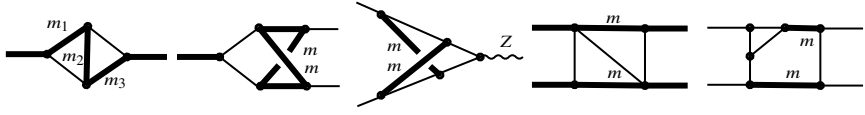
\begin{figure}
    \centering
    $\adjustbox{valign=c,scale={1}{1}}{\input{tikz/fig1}}$
    \caption{Examples of graphs in which one (or more) elliptic curves is (are) lurking. These include self-energy kite integrals, electro-weak form-factors and Bhabha scatterings.}
    \label{fig:fig1}
\end{figure}

For diagrams with loops, the corresponding \emph{Feynman integrals} over the off-shell loop momenta determine the probabilities of various particle interactions. Beyond the 1-loop level, Feynman integrals depending on multiple kinematic scales (like energies, angles, and masses) often involve complex underlying geometric structures. Fig.~\ref{fig:fig1} shows a few phenomenologically relevant 2-loop examples that involve one or more elliptic curves. 
As our understanding of these integrals deepened, it became evident that many could be totally understood and even evaluated through inherent properties of these rich geometries.

An effective approach employed to accomplish this is the method of \emph{differential equations}. However, a significant hurdle in using this approach is the path to a so-called \emph{canonical basis}. Such a basis, denoted by $\Vec{I}$, satisfies a differential equation that we can solve order-by-order in the dimensional regularization parameter $\varepsilon\in\mathbb{C}$ \cite{Henn:2013pwa}. It typically takes the following form:
\begin{equation}
\text{d}\Vec{I}=\varepsilon~\boldsymbol{\Omega}\cdot \Vec{I}\,,
\end{equation}
where $\boldsymbol{\Omega}$ is a matrix of kinematic one forms independent of $\varepsilon$. For many integrals (e.g., those without internal masses), a canonical basis is systematically obtained normalizing by \emph{leading singularities} (in the polylogarightmic case, one can think of leading singularities simply as maximal residues or as powers of Gram determinants in the external kinematics.) 

As soon as we turn on some internal masses, the story often changes. Here, we argue that this is because the precise definition of leading singularities in some of these examples is somewhat blurry (i.e., not uniquely defined), and so are \say{systematic methods} used to achieve a canonical form. In that sense, one of the most pressing problems in this program is how to construct such a basis $\Vec{I}$ from first principles, i.e., without relying too heavily on \emph{ad-hoc} ansatz-based techniques. Interesting progress on this issue was made recently in \cite{Gorges:2023zgv}.

While we do not claim to solve this problem fully, we believe that we can make significant progress using a concise set of analytical tools: unitarity and geometry. These tools are intrinsically integrated into the \emph{dual forms} framework of \cite{Caron-Huot:2021xqj,Caron-Huot:2021iev}. For this reason, dual forms seems to be the appropriate objects to leverage the intuitive, yet subtle, notion that a multi-loop problem is simply a bunch of coupled 1-loop problems \cite{Giroux:2022wav}. Furthermore, when working with elliptic Feynman integrals, it is to our advantage to foreground a key symmetry: the \emph{modular symmetry}. When combined into one toolbox, the above provides streamlined instructions to break down the construction of a canonical basis into a minimal sequence of algebraic steps. Below, we will demonstrate this approach using a proof-of-concept example: the unequal mass elliptic sunrise.

\section{From Feynman integrals to the dual paradigm}\label{sec:FI}
To set the stage, let us revisit the definition of Feynman integrals within dimensional regularization. At the most basic level, these are the integrals we obtain by applying Feynman rules on loop diagrams. However, as is the case with any other function, there isn't a unique way to represent it as an integral. Over the years, various representations have been introduced, each tailored for a specific purpose (for an extensive pedagogical account, see \cite{Weinzierl:2022eaz}.) Most representations take the form of \emph{twisted periods}:
\begin{equation}\label{eq:twistedPeriod}
    I=\int_X u~\phi\,,
\end{equation}
where the \emph{twist} $u$ typically represents a multi-valued function through its dependence on the (non-integer) spacetime dimension. Meanwhile, $X$ denotes an integration domain that incorporates the Feynman $i0$ prescription. Both $u$ and $X$ are common to all Feynman integrals in a given family, whereas the algebraic differential $n$-form $\phi$ is not shared among all integrals, and its form depends on the propagator structure of each graph. Thus, the purpose behind expressing Feynman integrals as shown in \eqref{eq:twistedPeriod} is to highlight the distinction between quantities such as $u$ and $X$ that are universal to a family of integrals, and those tied to a particular diagram such as $\phi$.

An important property of Feynman integrals is that they can consistently be represented as a \emph{finite} combination of simpler integrals \cite{Smirnov:2010hn}. This finite collection of integrals is commonly referred to as the set of \emph{master integrals}. It follows that any family of Feynman integrals forms a vector space. This vector space also possesses the additional property of being closed under differentiation with respect to the kinematic variables \cite{Frellesvig:2019uqt}.

Now, one can invoke a standard fact from linear algebra: for any given finite-dimensional vector space, there exists a \emph{dual vector space}, from which the original space can be studied.  Our next objective is to understand how the dual space for Feynman integrals is defined and its practical use in computations.
\paragraph{Dual forms}
In the most direct manner, the integrand 
$\widecheck{\phi}$ dual to the Feynman integrand 
$\phi$ (as given in \eqref{eq:twistedPeriod}) is defined such that the \emph{intersection paring}:
\begin{equation}\label{eq:defIN}
   \langle \widecheck{\phi}|\phi\rangle\propto\int_{\Gamma} ~(\widecheck{u}\times u)~\widecheck{\phi}{{\wedge}}\phi\,,
\end{equation}
makes sense.\footnote{See \cite{Caron-Huot:2021iev} for the exact constant of proportionality.} Here, $\Gamma\equiv\mathbb{C}^{n}\setminus\{u=0\}\cup\{\text{propagators}=0\}$ and $\widecheck{u}\equiv u|_{\varepsilon\to -\varepsilon}$ denotes the \emph{dual twist}. 

Let us unpack this further. For the intersection number \eqref{eq:defIN} to be meaningful, it must be unique. This uniqueness requires that the product of the dual twist and the twist is an algebraic function (not necessarily one!) Furthermore, we need this unique number to be finite. This condition is met when the zeros of the dual form are paired with the poles of the Feynman integrands, which are the propagators. In other words, $\widecheck{\phi}$ is supported only away from $\phi$'s \emph{unregulated} (\emph{non-twisted}) poles.

In practice, we implement this condition by attaching strings of  $\delta$-functions to dual forms for each propagator on the Feynman side: the form $\widecheck{\phi}$ dual to the Feynman form $\phi$ is proportional to a wedge product of $\text{d}\theta(x)=\delta(x)\text{d}x$'s (in the sense of distributions), each of which accounts for a propagator that can be put on-shell. The overall function multiplying this wedge product varies depending on the situation, as we will see later. For a complete treatment of dual forms, see \cite{Caron-Huot:2021xqj,Caron-Huot:2021iev,Giroux:2022wav}.

\section{Looping the loops and differential equations}\label{sec:lbl}

Now that the basics of dual forms are established, let us describe in detail the method used below to generate differential equations for dual Feynman integrals. This method draws from the naive idea that a multi-loop problem can be broken down into a collection of simpler, yet coupled, 1-loop problems. We will see below how this can be used to construct differential equations for (dual) Feynman integrals, one loop at a time.  This approach is interesting from the standpoint of multi-scales problems because it comes with the immediate reward of having to deal with less variables at a time.

\paragraph{Setup}

To turn a multi-loop problem into a series of more manageable 1-loop problems, it is essential to get a grip on the fibre bundle structure of the multi-loop integrals we want to study. 
Given that a 2-loop integrand effectively illustrates this concept and that the pattern is evident for $(L>2)$-loop integrands, we focus on 2-loop integrals.

Denoting by $\ell_1$ and $\ell_2$ the loop momenta, one may rewrite any 2-loop integrand as the wedge product of some (fibered) 1-loop basis:
\begin{align}\label{eq:lblSplit}
    \underbracket[0.4pt]{\widecheck{\varphi}^{\text{\tiny(2-loop)}}_a(\ell_1, \ell_2, \{p_{ij}^2\},\{m_j\})}_{\text{2-loop integrand (total space)}}
    = \underbracket[0.4pt]{\widecheck{\varphi}^{\text{\tiny(1-loop)}}_b(\ell_2, \{q_{ij}^2\},\{m_j\})}_{\text{First loop (fibre)}} 
    {{\wedge}} \underbracket[0.4pt]{\widecheck{\varphi}^{\text{\tiny(left-over)}}_{ba}(\ell_1, \{p_{ij}^2\},\{m_j\})}_{\text{Second loop (base)}}\,,
\end{align}
with $p_{ij}=p_i{+}p_j$, where the $p$'s denote the external momenta in the 2-loop integrand (total space). Similarly, $q_{ij}=q_i{+}q_j$, where the $q$'s label the external momenta in the first loop integrand (fibre), which can include some of the $p$'s as well as $\ell_1$.

The splitting given in \eqref{eq:lblSplit} imposes constraints on the left-over pieces (base) and guides our selection of ``good'' $\widecheck{\varphi}^{\text{\tiny(left-over)}}_{ba}$. In addition, once the differential equation $\boldsymbol{\Omega}^{\text{\tiny(1-loop)}}$ for the 1-loop basis is known, we can commute $\widecheck{\nabla}\equiv\text{d}+\text{d}\log \widecheck{u}^{\text{\tiny(1-loop)}}{{\wedge}}$ across the 1-loop basis to get a new covariant derivative acting on the left-over part: 
\begin{align}\label{eq:step1}
    \widecheck{\nabla} \left( \widecheck{\varphi}^{\text{\tiny(1-loop)}}_b
    {{\wedge}} \widecheck{\varphi}^{\text{\tiny(left-over)}}_{ba} \right)
    \simeq \widecheck{\varphi}^{\text{\tiny(1-loop)}}_b {{\wedge}} \widecheck{\nabla}_{bc}^{\text{\tiny(new)}}
    \left(\widecheck{\varphi}^{\text{\tiny(left-over)}}_{ca} \right)\,. 
\end{align}

Above,\say{$\simeq$} stands for \say{equal mod total derivatives.} Moreover, the knowing $\boldsymbol{\Omega}^{\text{\tiny(1-loop)}}$ gives us the covariant derivative $\widecheck{\nabla}_{ab}^{\text{\tiny(new)}}$ for the left-over part. Indeed, after the dust settles, one finds from \eqref{eq:step1}
\begin{align}\label{eq:baseCD}
\begin{split}
    \widecheck{\nabla}_{ab}^{\text{\tiny(new)}}&= \delta_{ab} \big(\textnormal{d}+\text{d}\log \widecheck{u}^{\text{\tiny(left-over)}}\big) + \Omega_{ab}^{\text{\tiny(1-loop)}} {{\wedge}}\\&\equiv \delta_{ab}\textnormal{d}+\widecheck{\omega}_{ab}{{\wedge}}
\end{split}
    \quad, \text{where} \quad 
     \widecheck{\nabla} \widecheck{\varphi}^{\text{\tiny(1-loop)}}_{a}
    \simeq \widecheck{\varphi}^{\text{\tiny(1-loop)}}_{b} 
    {{\wedge}} \Omega_{ba}^{\text{\tiny(1-loop)}}\,.
\end{align}

We expect that it should be easier to obtain a \say{good} (but not necessarily $\varepsilon$-form) 2-loop differential equations from the action of $\widecheck{\nabla}_{bc}^{\text{\tiny(new)}}$ on $\widecheck{\varphi}^{\text{\tiny(left-over)}}_{ca}$ than from the action of $\widecheck{\nabla}$ on $\widecheck{\varphi}^{\text{\tiny(2-loop)}}$:
\begin{equation}\label{eq:exp}
\text{\emph{Computing}}~~\widecheck{\nabla}\widecheck{\varphi}^{\text{\tiny(2-loop)}}_{a}\simeq \widecheck{\varphi}^{\text{\tiny(2-loop)}}_{b}{{\wedge}}\Omega_{ba}^{\text{\tiny(2-loop)}}~~\text{\emph{is harder than}}~~\widecheck{\nabla}_{bc}^{\text{\tiny(new)}}\widecheck{\varphi}^{\text{\tiny(left-over)}}_{ca}\simeq\widecheck{\varphi}^{\text{\tiny(left-over)}}_{bc}{{\wedge}}\Omega_{ca}^{\text{\tiny(2-loop)}}\,.
\end{equation}

The rationale for this expectation is that this approach yields two strong constraints on the fibered dual bases, such that there are not too many options to start the problem with. We refer to these constraints below as \emph{loop-by-loop constraints}. The \labelword{Word:fLBL}{first} imposes that the fibre basis is normalized such that $\widecheck{\boldsymbol{\omega}}$ in \eqref{eq:baseCD} factorizes linearly in $\varepsilon$ -- i.e., that $\widecheck{\boldsymbol{\omega}}\propto\varepsilon$.
The \labelword{Word:sLBL}{second} requires the wedge product in \eqref{eq:lblSplit} to be single-valued (algebraic). Note that when this constraint is applied, the fibre forms have \emph{already} been fixed by the first constraint. Therefore, only the base form basis is affected by this constraint.

While the expectation given in \eqref{eq:exp} is quite general, there is an important technical caveat we would like address here. Indeed, although the (dual) IBP reduction resulting in $\boldsymbol{\Omega}^{\text{\tiny(2-loop)}}$ works well using a \emph{global} set of coordinates in some of the examples we examined (e.g., Sec.~\ref{sec:sunrise}), we found instances (e.g., the 5-mass kite integral family) with obstructions to this global approach \cite{giroux:2023}. One way out of this is to use multiple local coordinate systems, which allow for a reduction in each sector, instead of relying solely on a global one. The results can then be pulled back to a global set of coordinates post reduction. More optimistically, in view of the recent progress discussed in \cite{hjalteTalk}, one might soon be able to bypass the cumbersome use of local coordinates and IBPs by employing directly cutting-edge intersection numbers algorithms.

\section{The 3-mass sunrise}\label{sec:sunrise}
Let us now exemplify the discussion above and consider the scalar 2-loop 3-mass elliptic sunrise. From the Feynman rules, we find (up to an overall normalization):
\begin{equation}\label{eq:sunrise}
    \adjustbox{valign=c,scale={1}{1},raise=0.33em}{\input{tikz/sunrise}}=\int\frac{\delta^\text{D}(\ell_3{-}\ell_1{+}\ell_2{-}p)}{(\ell_1^2{+}m_1^2{-}i0)(\ell_2^2{+}m_2^2{-}i0)(\ell_3^2{+}m_3^2{-}i0)}\prod_{a=1}^3\frac{\text{d}^\text{D}\ell_a}{i\pi^{\text{D}/2}} \qquad (\text{D}=4{-}2\varepsilon)\,.
\end{equation}
\paragraph{The \emph{elliptic} sunrise}
From now on, it will be convenient to work directly in momentum space and, in particular, to consider the cylindrical-like parameterization for the internal loop momenta:
\begin{equation}
\ell_{i}=\ell_{i\|}+\ell_{i\perp}\,, \quad \ell_{i\|}\cdot\ell_{i\perp}=0\,, \quad \ell_{1\|}=\textcolor{Maroon}{x}~p\,,\quad \ell_{2\|}=y~(\ell_1+p)\,, \qquad \text{where}~i=1,2\,.
\end{equation}
In these variables, the propagators (boundaries) are given by
\begin{gather}\label{eq:propList}
    \textsf{D}_{1} = \ell_{1\perp}^{2} + \textcolor{Maroon}{x}^2 p^2 + m_{1}^{2}\,, \qquad \textsf{D}_{2} 
    =\ell_{2\perp}^{2}+y^2~(\ell_{1\perp}^2+(1+\textcolor{Maroon}{x})^2p^2)+m_2^2\,,\\
        \textsf{D}_{3} 
    =\ell_{2\perp}^{2} + \left(y+1\right)^{2}\ell_{1\perp}^{2} + \left(1+y\right)^{2}\left(1+\textcolor{Maroon}{x}\right)^{2}p^{2}+m_{3}^{2}\,.\notag
\end{gather}
We are now in a good position to understand why the sunrise integral is referred to as \say{elliptic.} If we compute the maximal cut ($\textsf{D}_i^{-1}{\to}2\pi i\delta(\textsf{D}_i)$ \ $\forall~i$) of \eqref{eq:sunrise} in the critical dimension ($\varepsilon=0$), we obtain a 1-fold integral over the square root $Y(\textcolor{Maroon}{x})$ of the irreducible quartic polynomial:
\begin{equation}\label{eq:ellCurve}
    E(\mathbb{C}):Y^2-(\textcolor{Maroon}{x}-r_1)(\textcolor{Maroon}{x}-r_2)(\textcolor{Maroon}{x}-r_3)(\textcolor{Maroon}{x}-r_4)=0\,.
\end{equation}
where the roots $r_i$ are explicitly recorded in \cite[Eq. (4.32)]{Giroux:2022wav}. In mathematics, the object $E(\mathbb{C})$ given in \eqref{eq:ellCurve} is known as an \emph{elliptic curve} and has a rich and well-documented geometric structure. In particular, for a detailed discussion on its relationship with tori, refer to \cite[\S 4.3]{Giroux:2022wav}. 

\subsection{Looping the sunrise's loops: dual bases and differential equation}
The next step involves constructing an explicit 2-loop basis based on the loop-by-loop splitting presented in \eqref{eq:lblSplit}. This procedure is schematically summarized in Fig.~\ref{fig:fig2}.
\begin{figure}
    \centering
    $\adjustbox{valign=c,scale={1}{1}}{\input{tikz/fullSpaceBasis}}$
    $\adjustbox{valign=c,scale={1}{1}}{\input{tikz/lblBasis}}$
    \caption{The splitting in \eqref{eq:lblSplit} is schematically shown for the 2-loop sunrise. \textbf{Top:} The 2-loop (total space) dual basis, where both $\ell_1$ and $\ell_2$ are active integration variables.  \textbf{Middle:} The 1-loop (fibre) basis, where only $\ell_2$ is active. \textbf{Bottom:} The left-over (base) basis, where $\ell_2$ is integrated out already and $\ell_1$ is active.}
    \label{fig:fig2}
\end{figure}
We've divided the discussion into two steps, and further details can be found in \cite{Giroux:2022wav}.
\paragraph{Step I: fixing the 1-loop (fibre) basis} Examining the middle panel of Fig.~\eqref{fig:fig2}, we notice that the fibre basis is essentially a dual bubble basis (up to normalization). As established in \cite{Caron-Huot:2021xqj}, a canonical basis for the dual bubble, in the sense of \cite{Henn:2013pwa}, is
\begin{equation}
\widecheck{\varphi}_{1}^{\text{\tiny~(1-loop)}}=N \ \frac{2\varepsilon~\text{d}\theta_2{\wedge}{\text{d}}\ell_{2\parallel}}{\sqrt{(p{+}\ell_1)^2}\left.\ell_{2\perp}^{2}\right|_{2}}\,,~ 
\widecheck{\varphi}_{2}^{\text{\tiny~(1-loop)}}=N \frac{2\varepsilon~\text{d}\theta_3{\wedge}{\text{d}}\ell_{2\parallel}}{\sqrt{(p{+}\ell_1)^2}\left.\ell_{2\perp}^{2}\right|_{3}}\,,~ 
\widecheck{\varphi}_{3}^{\text{\tiny~(1-loop)}}=N\frac{\text{d}\theta_2{\wedge} \text{d}\theta_3}{\sqrt{(p{+}\ell_1)^2\left.\ell_{2\perp}^{2}\right|_{23}}}\,,
\end{equation}
where $N$ is a constant. Here, we have abbreviated $\theta(\text{D}_i)$ as $\theta_i$, with $\text{D}_i$ specified in \eqref{eq:propList}. 

Subsequently, the $\ref{Word:fLBL}^{\text{st}}$ loop-by-loop constraint instructs us to promote $N$ to a function of the kinematics. This constraint uniquely determines $N$ as the Gram determinant on the left-over loop -- i.e., $N=(\ell_{1\perp}^2)^{-1/2}$. Consequently, when localized on the boundary where all propagators are on-shell, the bubble form denominator becomes proportional to $Y$ as given in \eqref{eq:ellCurve}. This renders the fibre basis non-algebraic. Consequently, through the $\ref{Word:sLBL}^{\text{nd}}$ loop-by-loop constraint, further restrictions are imposed on the remaining (base) basis.

\paragraph{Step II: fixing the left-over (base) basis} 
In addition to the $\ref{Word:sLBL}^{\text{nd}}$ loop-by-loop constraint, one might want the base basis to exhibit certain properties that feel simply natural. For instance, we may ask the base basis to be as close as possible of being uniformly transcendental (readers with exposure to polylogarithmic Feynman integrals will likely find this assumption legitimate.) Furthermore, we can narrow our attention to bases satisfying a differential equation with a specific modular transformation rule: it must be independent of the modular parameters $a$ and $b$ under a modular transformation (defined in \cite[Eq. (4.52)]{Giroux:2022wav}.) Such a condition is crucial if one aims to rewrite/pullback the differential equation, initially written in terms of Mandelstam invariants and masses, into a modular form spanned by modular and Kronecker forms (a comprehensive review aimed at physicists can be found in \cite[\S 13]{Weinzierl:2022eaz}.)

The simplest basis that satisfies all these assumptions is presented in of \cite[Eqs. (5.2-5.3)]{Giroux:2022wav}. Notably, the elliptic sector is both surprisingly compact and natural:
\begin{equation}\label{eq:basis1}
    \big\{\textcolor{black}{\widecheck{\varphi} _{ij}^{\text{\tiny~(left-over)}}}\big\}_{j=4}^{7} =\left\{\frac{\psi_1^2}{\pi~ \varepsilon~W_{X}}\widecheck{\nabla}_{X}^{\text{\tiny(new)}},\frac{(x{-}r_1)\psi_1}{\pi},\frac{Y(c)\psi_1}{\pi(x-c)},1\right\}\frac{\pi\textnormal{d}\theta_1{\wedge}\textnormal{d}x}{m_1^{4\varepsilon}\textcolor{black}{\psi_1} Y}\begin{tiny}\begin{pmatrix}0\\0\\1\end{pmatrix}\end{tiny}\,.
\end{equation}

This basis is considered natural because the first and last forms span the cohomology of the bare (unpunctured) elliptic curve. The other two meromorphic forms have poles at twisted poles ($x=\infty$ and $x=c$, respectively) in $\text{D}=4$, which accurately accounts for the two punctures present on the given elliptic curve, as illustrated in \cite[Fig. 3]{Giroux:2022wav}. In the basis above, $X=\frac{p^2}{m_3^2}$, $\psi_1$ denotes one period of the elliptic curve and the term $W_X$ stands for a Wronskian. The various factors of $\pi$ and $\varepsilon$ ensure the basis is as close as possible of being of uniform transcendental weight. 

Interestingly, this basis possesses another property that we did not explicitly enforce: its differential equation is linear and strictly lower-triangular. In other words, it takes the form:
\begin{equation}
    \boldsymbol{\Omega}^{\text{\tiny(2-loop)}}=\boldsymbol{\Omega}^{\text{\tiny(2-loop)}}_{(0)}+\varepsilon~\boldsymbol{\Omega}^{\text{\tiny(2-loop)}}_{(1)}\,, \quad \text{with}~\boldsymbol{\Omega}^{\text{\tiny(2-loop)}}_{(1)}~\text{lower-triangular}\,. 
\end{equation}
Notice that $\boldsymbol{\Omega}^{\text{\tiny(2-loop)}}$ is \emph{not} in $\varepsilon$-form. To achieve this form, a gauge transformation $\boldsymbol{U}$ must be performed on \eqref{eq:basis1} such that the new differential equation reads
\begin{equation}
   \varepsilon~\tilde{\boldsymbol{\Omega}}^{\text{\tiny(2-loop)}}_{(1)}=(\boldsymbol{U}\cdot\boldsymbol{\Omega}^{\text{\tiny(2-loop)}}+\text{d}\boldsymbol{U})\cdot\boldsymbol{U}^{-1}\,.
\end{equation}
As highlighted in \cite{Giroux:2022wav} and further verified with more general examples in \cite{giroux:2023}, the matrix $\boldsymbol{U}$ is entirely determined by modular symmetry. This implies that the differential equation
\begin{equation}\label{eq:tosolve}
\boldsymbol{U}\cdot\boldsymbol{\Omega}^{\text{\tiny(2-loop)}}_{(0)}+\text{d}\boldsymbol{U}=0\,,
\end{equation}
can be solved analytically for $\boldsymbol{U}$ without performing any integration (!), simply by asking that \eqref{eq:tosolve} is modular covariant. The matrix $\boldsymbol{U}$ specific to the sunrise example is recorded in \cite[App. E]{Giroux:2022wav}. The corresponding differential equation is provided in \cite[Eq. (5.81)]{Giroux:2022wav} and is demonstrated to have a simple relation to that of Feynman integrals considered in \cite{Bogner:2019lfa}.

\section{Conclusion}\label{sec:conclusion}

Driven by the idea that multi-loop Feynman integrals are an iteration over simpler 1-loop problems, we developed a loop-by-loop method for the computation of multi-loop dual Feynman integrands' differential equations. Through the intersection number, dual Feynman integrands are in one-to-one correspondence with \say{conventional} Feynman integrands. These dual integrands, supported on generalized unitarity cuts, are inherently simpler.

In this talk, the primary benefit of working with dual integrands was the ability to localize to generalized unitarity cuts, which later on informed on an optimal choice of basis. The geometry tied to a Feynman integrand is often concealed within its cuts, and dual integrands bring this to the fore. Moreover, because dual integrands aren't constrained to \say{look like} traditional Feynman integrands, we considered a loop-by-loop basis that is entirely motivated from the underlying elliptic geometry.

As a simple yet non-trivial example, we constructed an $\varepsilon$-form dual basis for the three mass 2-loop elliptic sunrise family and the associated differential equation. In doing so, we saw that breaking up the problem into simpler 1-loop problems yields several advantages. First, we can reuse the known $\varepsilon$-form basis and differential equation at 1-loop to construct the 2-loop basis and differential equation. Second, at each step only a small subset of variables are active on the fibre simplifying algebraic manipulations as well as integration by-parts.

While the starting base basis \eqref{eq:basis1} was geometrically and physically well motivated, it did not have an  $\varepsilon$-form differential equation. A gauge transformation was needed to bring the system into $\varepsilon$-form, and we saw that it was entirely fixed by modular covariance. We anticipate that this idea will prove useful for future cutting-edge computations of differential equations for topologies with elliptic-like geometries, whether one uses the dual form framework or not.

\paragraph{Acknowldgments}\label{sec:conclusion}
This work was co-funded by the National Science and Engineering Council of Canada (NSERC) (MG), the Simons Investigator Award $\#$376208 (AP), and the European Union (ERC Consolidator Grant LoCoMotive 101043686 (FP)). 

\bibliographystyle{JHEP}
\bibliography{main}
\end{document}

%% file: tikz/fig1.tex
\tikzset{every picture/.style={line width=0.75pt}}         
\begin{tikzpicture}[x=0.75pt,y=0.75pt,yscale=-0.7,xscale=0.7]

\draw [line width=2.25]    (3,51) -- (33.43,51) ;
\draw [line width=2.25]    (32.43,51) -- (61.43,31) ;
\draw [shift={(61.43,31)}, rotate = 325.41] [color={rgb, 255:red, 0; green, 0; blue, 0 }  ][fill={rgb, 255:red, 0; green, 0; blue, 0 }  ][line width=2.25]      (0, 0) circle [x radius= 1, y radius= 1]   ;
\draw [shift={(32.43,51)}, rotate = 325.41] [color={rgb, 255:red, 0; green, 0; blue, 0 }  ][fill={rgb, 255:red, 0; green, 0; blue, 0 }  ][line width=2.25]      (0, 0) circle [x radius= 1, y radius= 1]   ;
\draw [line width=2.25]    (61.43,31) -- (60.43,73) ;
\draw    (32.43,51) -- (60.43,73) ;
\draw    (61.43,31) -- (89.43,53) ;
\draw [line width=2.25]    (60.43,73) -- (89.43,53) ;
\draw [shift={(89.43,53)}, rotate = 325.41] [color={rgb, 255:red, 0; green, 0; blue, 0 }  ][fill={rgb, 255:red, 0; green, 0; blue, 0 }  ][line width=2.25]      (0, 0) circle [x radius= 1, y radius= 1]   ;
\draw [shift={(60.43,73)}, rotate = 325.41] [color={rgb, 255:red, 0; green, 0; blue, 0 }  ][fill={rgb, 255:red, 0; green, 0; blue, 0 }  ][line width=2.25]      (0, 0) circle [x radius= 1, y radius= 1]   ;
\draw [line width=2.25]    (87.43,53) -- (117.86,53) ;
\draw [line width=2.25]    (125,53) -- (155.43,53) ;
\draw [shift={(155.43,53)}, rotate = 360] [color={rgb, 255:red, 0; green, 0; blue, 0 }  ][fill={rgb, 255:red, 0; green, 0; blue, 0 }  ][line width=2.25]      (0, 0) circle [x radius= 1, y radius= 1]   ;
\draw    (155.43,53) -- (183.43,75) ;
\draw    (183.43,32) -- (155.43,53) ;
\draw [line width=2.25]    (183.43,32) -- (217.43,74) ;
\draw [shift={(217.43,74)}, rotate = 51.01] [color={rgb, 255:red, 0; green, 0; blue, 0 }  ][fill={rgb, 255:red, 0; green, 0; blue, 0 }  ][line width=2.25]      (0, 0) circle [x radius= 1, y radius= 1]   ;
\draw [shift={(183.43,32)}, rotate = 51.01] [color={rgb, 255:red, 0; green, 0; blue, 0 }  ][fill={rgb, 255:red, 0; green, 0; blue, 0 }  ][line width=2.25]      (0, 0) circle [x radius= 1, y radius= 1]   ;
\draw [line width=2.25]    (183.43,75) -- (197.57,59) ;
\draw [line width=2.25]    (205,49) -- (216.43,32) ;
\draw [line width=2.25]    (183.43,33) -- (216.43,33) ;
\draw [shift={(216.43,33)}, rotate = 0] [color={rgb, 255:red, 0; green, 0; blue, 0 }  ][fill={rgb, 255:red, 0; green, 0; blue, 0 }  ][line width=2.25]      (0, 0) circle [x radius= 1, y radius= 1]   ;
\draw [line width=2.25]    (184.43,74) -- (217.43,74) ;
\draw [shift={(184.43,74)}, rotate = 0] [color={rgb, 255:red, 0; green, 0; blue, 0 }  ][fill={rgb, 255:red, 0; green, 0; blue, 0 }  ][line width=2.25]      (0, 0) circle [x radius= 1, y radius= 1]   ;
\draw    (217.43,74) -- (241,74) ;
\draw    (216.43,33) -- (240,33) ;
\draw [line width=0.75]    (371.56,52.98) .. controls (369.9,54.65) and (368.23,54.66) .. (366.56,52.99) .. controls (364.89,51.33) and (363.22,51.34) .. (361.56,53.01) .. controls (359.89,54.68) and (358.23,54.68) .. (356.56,53.02) .. controls (354.89,51.36) and (353.23,51.36) .. (351.56,53.03) .. controls (349.89,54.7) and (348.23,54.7) .. (346.56,53.04) .. controls (344.89,51.38) and (343.22,51.39) .. (341.56,53.06) -- (341.14,53.06) -- (341.14,53.06) ;
\draw [shift={(341.14,53.06)}, rotate = 179.86] [color={rgb, 255:red, 0; green, 0; blue, 0 }  ][fill={rgb, 255:red, 0; green, 0; blue, 0 }  ][line width=2.25]      (0, 0) circle [x radius= 1, y radius= 1]   ;
\draw    (341.14,53.06) -- (252.86,13.86) ;
\draw    (253.57,88.43) -- (341.14,53.06) ;
\draw [line width=2.25]    (270.43,22.57) -- (297.43,50) ;
\draw [shift={(270.43,22.57)}, rotate = 45.45] [color={rgb, 255:red, 0; green, 0; blue, 0 }  ][fill={rgb, 255:red, 0; green, 0; blue, 0 }  ][line width=2.25]      (0, 0) circle [x radius= 1, y radius= 1]   ;
\draw [line width=2.25]    (317.57,42.43) -- (270.43,80.57) ;
\draw [shift={(270.43,80.57)}, rotate = 141.02] [color={rgb, 255:red, 0; green, 0; blue, 0 }  ][fill={rgb, 255:red, 0; green, 0; blue, 0 }  ][line width=2.25]      (0, 0) circle [x radius= 1, y radius= 1]   ;
\draw [shift={(317.57,42.43)}, rotate = 141.02] [color={rgb, 255:red, 0; green, 0; blue, 0 }  ][fill={rgb, 255:red, 0; green, 0; blue, 0 }  ][line width=2.25]      (0, 0) circle [x radius= 1, y radius= 1]   ;
\draw [line width=2.25]    (378,32) -- (408.43,32) ;
\draw [shift={(408.43,32)}, rotate = 360] [color={rgb, 255:red, 0; green, 0; blue, 0 }  ][fill={rgb, 255:red, 0; green, 0; blue, 0 }  ][line width=2.25]      (0, 0) circle [x radius= 1, y radius= 1]   ;
\draw [line width=2.25]    (378,73) -- (408.43,73) ;
\draw [shift={(408.43,73)}, rotate = 360] [color={rgb, 255:red, 0; green, 0; blue, 0 }  ][fill={rgb, 255:red, 0; green, 0; blue, 0 }  ][line width=2.25]      (0, 0) circle [x radius= 1, y radius= 1]   ;
\draw    (408.43,32) -- (408.43,73) ;
\draw [line width=2.25]    (408.43,32) -- (458.57,32.57) ;
\draw [shift={(458.57,32.57)}, rotate = 0.65] [color={rgb, 255:red, 0; green, 0; blue, 0 }  ][fill={rgb, 255:red, 0; green, 0; blue, 0 }  ][line width=2.25]      (0, 0) circle [x radius= 1, y radius= 1]   ;
\draw [line width=2.25]    (408.43,73) -- (458.57,73.57) ;
\draw [shift={(458.57,73.57)}, rotate = 0.65] [color={rgb, 255:red, 0; green, 0; blue, 0 }  ][fill={rgb, 255:red, 0; green, 0; blue, 0 }  ][line width=2.25]      (0, 0) circle [x radius= 1, y radius= 1]   ;
\draw    (458.57,32.57) -- (458.57,73.57) ;
\draw    (408.43,32) -- (458.57,73.57) ;
\draw [line width=2.25]    (458.57,32.57) -- (489,32.57) ;
\draw [line width=2.25]    (458.57,73.57) -- (489,73.57) ;
\draw [line width=0.75]    (503,32) -- (533.43,32) ;
\draw [shift={(533.43,32)}, rotate = 360] [color={rgb, 255:red, 0; green, 0; blue, 0 }  ][fill={rgb, 255:red, 0; green, 0; blue, 0 }  ][line width=2.25]      (0, 0) circle [x radius= 1, y radius= 1]   ;
\draw [line width=0.75]    (503,73) -- (533.43,73) ;
\draw [shift={(533.43,73)}, rotate = 360] [color={rgb, 255:red, 0; green, 0; blue, 0 }  ][fill={rgb, 255:red, 0; green, 0; blue, 0 }  ][line width=2.25]      (0, 0) circle [x radius= 1, y radius= 1]   ;
\draw    (533.43,32) -- (533.43,73) ;
\draw [line width=2.25]    (558.43,32) -- (583.57,32.57) ;
\draw [shift={(583.57,32.57)}, rotate = 1.3] [color={rgb, 255:red, 0; green, 0; blue, 0 }  ][fill={rgb, 255:red, 0; green, 0; blue, 0 }  ][line width=2.25]      (0, 0) circle [x radius= 1, y radius= 1]   ;
\draw [shift={(558.43,32)}, rotate = 1.3] [color={rgb, 255:red, 0; green, 0; blue, 0 }  ][fill={rgb, 255:red, 0; green, 0; blue, 0 }  ][line width=2.25]      (0, 0) circle [x radius= 1, y radius= 1]   ;
\draw [line width=2.25]    (533.43,73) -- (583.57,73.57) ;
\draw [shift={(583.57,73.57)}, rotate = 0.65] [color={rgb, 255:red, 0; green, 0; blue, 0 }  ][fill={rgb, 255:red, 0; green, 0; blue, 0 }  ][line width=2.25]      (0, 0) circle [x radius= 1, y radius= 1]   ;
\draw    (583.57,32.57) -- (583.57,73.57) ;
\draw    (533.43,52.5) -- (558.43,32) ;
\draw [shift={(533.43,52.5)}, rotate = 320.65] [color={rgb, 255:red, 0; green, 0; blue, 0 }  ][fill={rgb, 255:red, 0; green, 0; blue, 0 }  ][line width=2.25]      (0, 0) circle [x radius= 1, y radius= 1]   ;
\draw [line width=0.75]    (583.57,32.57) -- (614,32.57) ;
\draw [line width=0.75]    (583.57,73.57) -- (614,73.57) ;
\draw    (533.43,32) -- (558.43,32) ;
\draw [line width=2.25]    (305.43,59.71) -- (312.43,66) ;
\draw [shift={(312.43,66)}, rotate = 41.92] [color={rgb, 255:red, 0; green, 0; blue, 0 }  ][fill={rgb, 255:red, 0; green, 0; blue, 0 }  ][line width=2.25]      (0, 0) circle [x radius= 1, y radius= 1]   ;

\draw (30,24.4) node [anchor=north west][inner sep=0.75pt]  [font=\tiny]  {$m_{1}$};
\draw (40.93,48.4) node [anchor=north west][inner sep=0.75pt]  [font=\tiny]  {$m_{2}$};
\draw (70.93,65.4) node [anchor=north west][inner sep=0.75pt]  [font=\tiny]  {$m_{3}$};
\draw (212.71,43.9) node [anchor=north west][inner sep=0.75pt]  [font=\tiny]  {$m$};
\draw (213.71,56.9) node [anchor=north west][inner sep=0.75pt]  [font=\tiny]  {$m$};
\draw (272.71,40.9) node [anchor=north west][inner sep=0.75pt]  [font=\tiny]  {$m$};
\draw (272.71,55.9) node [anchor=north west][inner sep=0.75pt]  [font=\tiny]  {$m$};
\draw (353,35.4) node [anchor=north west][inner sep=0.75pt]  [font=\tiny]  {$Z$};
\draw (428.71,20.9) node [anchor=north west][inner sep=0.75pt]  [font=\tiny]  {$m$};
\draw (429.71,60.9) node [anchor=north west][inner sep=0.75pt]  [font=\tiny]  {$m$};
\draw (556.71,60.9) node [anchor=north west][inner sep=0.75pt]  [font=\tiny]  {$m$};
\draw (566.71,37.9) node [anchor=north west][inner sep=0.75pt]  [font=\tiny]  {$m$};
\end{tikzpicture}

%% file: tikz/sunrise.tex
\tikzset{every picture/.style={line width=0.75pt}}
\begin{tikzpicture}
[x=0.75pt,y=0.75pt,yscale=-0.30,xscale=0.30]

\tikzset{ma/.style={decoration={markings,mark=at position 0.1 with {\arrow[scale=0.7]{>}}},postaction={decorate}}}
\tikzset{mar/.style={decoration={markings,mark=at position 0.5 with {\arrowreversed[scale=0.7]{>}}},postaction={decorate}}}

\draw   (260,128.81) .. controls (260,97.43) and (285.43,72) .. (316.81,72) .. controls (348.18,72) and (373.61,97.43) .. (373.61,128.81) .. controls (373.61,160.18) and (348.18,185.61) .. (316.81,185.61) .. controls (285.43,185.61) and (260,160.18) .. (260,128.81) -- cycle ;
\draw[ma]  (231,130) -- (401.8,130.2) ;
\draw (230.6,90.4) node [anchor=north west][inner sep=0.75pt]  [font=\scriptsize]  {$p$};
\draw (300.6,42.4) node [anchor=north west][inner sep=0.75pt]  [font=\scriptsize]  {$m_{1}$};
\draw (300.6,94.4) node [anchor=north west][inner sep=0.75pt]  [font=\scriptsize]  {$m_{2}$};
\draw (300.6,146.4) node [anchor=north west][inner sep=0.75pt]  [font=\scriptsize]  {$m_{3}$};
\end{tikzpicture}

%% file: tikz/fullSpaceBasis.tex
\tikzset{test/.style={
    postaction={
        decorate,
        decoration={
            markings,
            mark=at position \pgfdecoratedpathlength-0.5pt with {\arrow[RoyalBlue,line width=#1] {}; },
            mark=between positions 0 and \pgfdecoratedpathlength-5pt step 0.5pt with {
                \pgfmathsetmacro\myval{multiply(divide(
                    \pgfkeysvalueof{/pgf/decoration/mark info/distance from start}, \pgfdecoratedpathlength),100)};
                \pgfsetfillcolor{RoyalBlue!\myval!Maroon};
                \pgfpathcircle{\pgfpointorigin}{#1};
                \pgfusepath{fill};}
}}}}
\tikzset{every picture/.style={line width=0.75pt}}      
\begin{tikzpicture}[x=0.75pt,y=0.75pt,yscale=-0.8,xscale=0.8]

\draw    (101,80.5) -- (137.86,81.07) node[right]{\tiny$,$};
\draw  [color=RoyalBlue  ,draw opacity=1 ] (111.19,72.36) .. controls (111.19,72.36) and (111.19,72.36) .. (111.19,72.36) .. controls (111.19,72.36) and (111.19,72.36) .. (111.19,72.36) .. controls (106.64,67.71) and (106.72,60.25) .. (111.37,55.69) .. controls (116.02,51.14) and (123.48,51.22) .. (128.04,55.88) .. controls (132.59,60.53) and (132.51,67.99) .. (127.85,72.54) .. controls (122.24,78.04) and (119.43,80.79) .. (119.43,80.79) .. controls (119.43,80.79) and (116.68,77.98) .. (111.19,72.36) -- cycle ;
\draw  [color=Maroon  ,draw opacity=1 ] (128.06,88.81) .. controls (128.06,88.81) and (128.06,88.81) .. (128.06,88.81) .. controls (128.06,88.81) and (128.06,88.81) .. (128.06,88.81) .. controls (132.83,93.24) and (133.11,100.69) .. (128.68,105.46) .. controls (124.25,110.23) and (116.79,110.51) .. (112.02,106.08) .. controls (107.25,101.65) and (106.98,94.19) .. (111.41,89.42) .. controls (116.76,83.67) and (119.43,80.79) .. (119.43,80.79) .. controls (119.43,80.79) and (122.3,83.45) .. (128.06,88.81) -- cycle ;
\begin{scope}[xshift=-4pt]
\draw  [fill=black  ,fill opacity=1 ] (99.86,115.21) .. controls (96.25,115.21) and (93.32,98.95) .. (93.32,78.89) .. controls (93.32,58.83) and (96.25,42.57) .. (99.86,42.57) .. controls (96.62,44.78) and (94.11,60.2) .. (94.11,78.89) .. controls (94.11,97.59) and (96.62,113.01) .. (99.86,115.21) -- cycle ;
\end{scope}
\draw    (151.2,79.6) -- (188.86,80.07) node[right]{\tiny$,$};
\draw  [color=RoyalBlue  ,draw opacity=1 ] (162.19,71.36) .. controls (162.19,71.36) and (162.19,71.36) .. (162.19,71.36) .. controls (162.19,71.36) and (162.19,71.36) .. (162.19,71.36) .. controls (157.64,66.71) and (157.72,59.25) .. (162.37,54.69) .. controls (167.02,50.14) and (174.48,50.22) .. (179.04,54.88) .. controls (183.59,59.53) and (183.51,66.99) .. (178.85,71.54) .. controls (173.24,77.04) and (170.43,79.79) .. (170.43,79.79) .. controls (170.43,79.79) and (167.68,76.98) .. (162.19,71.36) -- cycle ;
\draw  [color=Maroon  ,draw opacity=1 ] (179.06,87.81) .. controls (179.06,87.81) and (179.06,87.81) .. (179.06,87.81) .. controls (179.06,87.81) and (179.06,87.81) .. (179.06,87.81) .. controls (183.83,92.24) and (184.11,99.69) .. (179.68,104.46) .. controls (175.25,109.23) and (167.79,109.51) .. (163.02,105.08) .. controls (158.25,100.65) and (157.98,93.19) .. (162.41,88.42) .. controls (167.76,82.67) and (170.43,79.79) .. (170.43,79.79) .. controls (170.43,79.79) and (173.3,82.45) .. (179.06,87.81) -- cycle ;
\draw    (201,79.5) -- (237.86,80.07) node[right]{\tiny$,$};
\draw  [color=RoyalBlue  ,draw opacity=1 ] (211.19,71.36) .. controls (211.19,71.36) and (211.19,71.36) .. (211.19,71.36) .. controls (211.19,71.36) and (211.19,71.36) .. (211.19,71.36) .. controls (206.64,66.71) and (206.72,59.25) .. (211.37,54.69) .. controls (216.02,50.14) and (223.48,50.22) .. (228.04,54.88) .. controls (232.59,59.53) and (232.51,66.99) .. (227.85,71.54) .. controls (222.24,77.04) and (219.43,79.79) .. (219.43,79.79) .. controls (219.43,79.79) and (216.68,76.98) .. (211.19,71.36) -- cycle ;
\draw  [color=Maroon  ,draw opacity=1 ] (228.06,87.81) .. controls (228.06,87.81) and (228.06,87.81) .. (228.06,87.81) .. controls (228.06,87.81) and (228.06,87.81) .. (228.06,87.81) .. controls (232.83,92.24) and (233.11,99.69) .. (228.68,104.46) .. controls (224.25,109.23) and (216.79,109.51) .. (212.02,105.08) .. controls (207.25,100.65) and (206.98,93.19) .. (211.41,88.42) .. controls (216.76,82.67) and (219.43,79.79) .. (219.43,79.79) .. controls (219.43,79.79) and (222.3,82.45) .. (228.06,87.81) -- cycle ;
\begin{scope}[xshift=4pt]
\draw  [fill=black  ,fill opacity=1 ] (546.32,41.57) .. controls (549.93,41.57) and (552.86,57.83) .. (552.86,77.89) .. controls (552.86,97.95) and (549.93,114.21) .. (546.32,114.21) .. controls (549.56,112.01) and (552.06,96.59) .. (552.06,77.89) .. controls (552.06,59.2) and (549.56,43.78) .. (546.32,41.57) -- cycle ;
\end{scope}
\draw  [draw opacity=0] (260.14,79.4) .. controls (260.14,79.4) and (260.14,79.4) .. (260.14,79.4) .. controls (260.14,66.76) and (269.42,56.51) .. (280.87,56.51) .. controls (292.23,56.51) and (301.46,66.6) .. (301.6,79.11) -- (280.87,79.4) -- cycle ; \draw  [color=RoyalBlue  ,draw opacity=1 ] (260.14,79.4) .. controls (260.14,79.4) and (260.14,79.4) .. (260.14,79.4) .. controls (260.14,66.76) and (269.42,56.51) .. (280.87,56.51) .. controls (292.23,56.51) and (301.46,66.6) .. (301.6,79.11) ;  
\draw  [draw opacity=0] (260.14,79.4) .. controls (260.14,79.4) and (260.14,79.4) .. (260.14,79.4) .. controls (260.14,90.81) and (269.42,100.05) .. (280.87,100.05) .. controls (292.22,100.05) and (301.44,90.96) .. (301.6,79.69) -- (280.87,79.4) -- cycle ; \draw  [color=Maroon  ,draw opacity=1 ] (260.14,79.4) .. controls (260.14,79.4) and (260.14,79.4) .. (260.14,79.4) .. controls (260.14,90.81) and (269.42,100.05) .. (280.87,100.05) .. controls (292.22,100.05) and (301.44,90.96) .. (301.6,79.69) ;  
\draw   (251,79.37) -- (310.74,79.43) node[right]{\tiny$,$};
\draw[test=0.6pt]    (260,79.37) -- (310.74,79.43) ;
\draw  [draw opacity=0] (335.14,79.4) .. controls (335.14,79.4) and (335.14,79.4) .. (335.14,79.4) .. controls (335.14,66.76) and (344.42,56.51) .. (355.87,56.51) .. controls (367.23,56.51) and (376.46,66.6) .. (376.6,79.11) -- (355.87,79.4) -- cycle ; \draw  [color=RoyalBlue  ,draw opacity=1 ] (335.14,79.4) .. controls (335.14,79.4) and (335.14,79.4) .. (335.14,79.4) .. controls (335.14,66.76) and (344.42,56.51) .. (355.87,56.51) .. controls (367.23,56.51) and (376.46,66.6) .. (376.6,79.11) ;  
\draw  [draw opacity=0] (335.14,79.4) .. controls (335.14,79.4) and (335.14,79.4) .. (335.14,79.4) .. controls (335.14,90.81) and (344.42,100.05) .. (355.87,100.05) .. controls (367.22,100.05) and (376.44,90.96) .. (376.6,79.69) -- (355.87,79.4) -- cycle ; \draw  [color=Maroon  ,draw opacity=1 ] (335.14,79.4) .. controls (335.14,79.4) and (335.14,79.4) .. (335.14,79.4) .. controls (335.14,90.81) and (344.42,100.05) .. (355.87,100.05) .. controls (367.22,100.05) and (376.44,90.96) .. (376.6,79.69) ;  
\draw   (326,79.37) -- (385.74,79.43) node[right]{\tiny$,$};
\draw[test=0.6pt]     (337,79.37) -- (386.74,79.43) ;
\draw  [draw opacity=0] (412.14,78.4) .. controls (412.14,78.4) and (412.14,78.4) .. (412.14,78.4) .. controls (412.14,65.76) and (421.42,55.51) .. (432.87,55.51) .. controls (444.23,55.51) and (453.46,65.6) .. (453.6,78.11) -- (432.87,78.4) -- cycle ; \draw  [color=RoyalBlue  ,draw opacity=1 ] (412.14,78.4) .. controls (412.14,78.4) and (412.14,78.4) .. (412.14,78.4) .. controls (412.14,65.76) and (421.42,55.51) .. (432.87,55.51) .. controls (444.23,55.51) and (453.46,65.6) .. (453.6,78.11) ;  
\draw  [draw opacity=0] (412.14,78.4) .. controls (412.14,78.4) and (412.14,78.4) .. (412.14,78.4) .. controls (412.14,89.81) and (421.42,99.05) .. (432.87,99.05) .. controls (444.22,99.05) and (453.44,89.96) .. (453.6,78.69) -- (432.87,78.4) -- cycle ; \draw  [color=Maroon  ,draw opacity=1 ] (412.14,78.4) .. controls (412.14,78.4) and (412.14,78.4) .. (412.14,78.4) .. controls (412.14,89.81) and (421.42,99.05) .. (432.87,99.05) .. controls (444.22,99.05) and (453.44,89.96) .. (453.6,78.69) ;  
\draw  (403,78.37) -- (462.74,78.43) node[right]{\tiny$,$};
\draw [test=0.6pt]     (411,78.37) -- (462.74,78.43) ;
\draw  [draw opacity=0] (492.14,77.4) .. controls (492.14,77.4) and (492.14,77.4) .. (492.14,77.4) .. controls (492.14,64.76) and (501.42,54.51) .. (512.87,54.51) .. controls (524.23,54.51) and (533.46,64.6) .. (533.6,77.11) -- (512.87,77.4) -- cycle ; \draw  [color=RoyalBlue  ,draw opacity=1 ] (492.14,77.4) .. controls (492.14,77.4) and (492.14,77.4) .. (492.14,77.4) .. controls (492.14,64.76) and (501.42,54.51) .. (512.87,54.51) .. controls (524.23,54.51) and (533.46,64.6) .. (533.6,77.11) ;  
\draw  [draw opacity=0] (492.14,77.4) .. controls (492.14,77.4) and (492.14,77.4) .. (492.14,77.4) .. controls (492.14,88.81) and (501.42,98.05) .. (512.87,98.05) .. controls (524.22,98.05) and (533.44,88.96) .. (533.6,77.69) -- (512.87,77.4) -- cycle ; \draw  [color=Maroon  ,draw opacity=1 ] (492.14,77.4) .. controls (492.14,77.4) and (492.14,77.4) .. (492.14,77.4) .. controls (492.14,88.81) and (501.42,98.05) .. (512.87,98.05) .. controls (524.22,98.05) and (533.44,88.96) .. (533.6,77.69) ;  
\draw   (483,77.37) -- (542.74,77.43) ;
\draw [test=0.6pt]    (489,77.37) -- (542.74,77.43) ;
\draw  [color=black  ,draw opacity=0 ][fill=black  ,fill opacity=1 ] (116.03,80.79) .. controls (116.03,78.91) and (117.55,77.39) .. (119.43,77.39) .. controls (121.31,77.39) and (122.83,78.91) .. (122.83,80.79) .. controls (122.83,82.66) and (121.31,84.19) .. (119.43,84.19) .. controls (117.55,84.19) and (116.03,82.66) .. (116.03,80.79) -- cycle ;
\draw  [color=black  ,draw opacity=0 ][fill=black  ,fill opacity=1 ] (166.63,79.84) .. controls (166.63,77.96) and (168.15,76.44) .. (170.03,76.44) .. controls (171.91,76.44) and (173.43,77.96) .. (173.43,79.84) .. controls (173.43,81.71) and (171.91,83.24) .. (170.03,83.24) .. controls (168.15,83.24) and (166.63,81.71) .. (166.63,79.84) -- cycle ;
\draw  [color=black  ,draw opacity=0 ][fill=black  ,fill opacity=1 ] (216.03,79.79) .. controls (216.03,77.91) and (217.55,76.39) .. (219.43,76.39) .. controls (221.31,76.39) and (222.83,77.91) .. (222.83,79.79) .. controls (222.83,81.66) and (221.31,83.19) .. (219.43,83.19) .. controls (217.55,83.19) and (216.03,81.66) .. (216.03,79.79) -- cycle ;
\draw  [color=black  ,draw opacity=0 ][fill=black  ,fill opacity=1 ] (256.74,79.4) .. controls (256.74,77.52) and (258.27,76) .. (260.14,76) .. controls (262.02,76) and (263.54,77.52) .. (263.54,79.4) .. controls (263.54,81.28) and (262.02,82.8) .. (260.14,82.8) .. controls (258.27,82.8) and (256.74,81.28) .. (256.74,79.4) -- cycle ;
\draw  [color=black  ,draw opacity=0 ][fill=black  ,fill opacity=1 ] (298.2,79.11) .. controls (298.2,77.23) and (299.72,75.71) .. (301.6,75.71) .. controls (303.48,75.71) and (305,77.23) .. (305,79.11) .. controls (305,80.98) and (303.48,82.51) .. (301.6,82.51) .. controls (299.72,82.51) and (298.2,80.98) .. (298.2,79.11) -- cycle ;
\draw  [color=black  ,draw opacity=0 ][fill=black  ,fill opacity=1 ] (331.74,79.4) .. controls (331.74,77.52) and (333.27,76) .. (335.14,76) .. controls (337.02,76) and (338.54,77.52) .. (338.54,79.4) .. controls (338.54,81.28) and (337.02,82.8) .. (335.14,82.8) .. controls (333.27,82.8) and (331.74,81.28) .. (331.74,79.4) -- cycle ;
\draw  [color=black  ,draw opacity=0 ][fill=black  ,fill opacity=1 ] (373.2,79.11) .. controls (373.2,77.23) and (374.72,75.71) .. (376.6,75.71) .. controls (378.48,75.71) and (380,77.23) .. (380,79.11) .. controls (380,80.98) and (378.48,82.51) .. (376.6,82.51) .. controls (374.72,82.51) and (373.2,80.98) .. (373.2,79.11) -- cycle ;
\draw  [color=black  ,draw opacity=0 ][fill=black  ,fill opacity=1 ] (408.74,78.4) .. controls (408.74,76.52) and (410.27,75) .. (412.14,75) .. controls (414.02,75) and (415.54,76.52) .. (415.54,78.4) .. controls (415.54,80.28) and (414.02,81.8) .. (412.14,81.8) .. controls (410.27,81.8) and (408.74,80.28) .. (408.74,78.4) -- cycle ;
\draw  [color=black  ,draw opacity=0 ][fill=black  ,fill opacity=1 ] (450.2,78.69) .. controls (450.2,76.81) and (451.72,75.29) .. (453.6,75.29) .. controls (455.47,75.29) and (457,76.81) .. (457,78.69) .. controls (457,80.57) and (455.47,82.09) .. (453.6,82.09) .. controls (451.72,82.09) and (450.2,80.57) .. (450.2,78.69) -- cycle ;
\draw  [color=black  ,draw opacity=0 ][fill=black  ,fill opacity=1 ] (488.74,77.4) .. controls (488.74,75.52) and (490.27,74) .. (492.14,74) .. controls (494.02,74) and (495.54,75.52) .. (495.54,77.4) .. controls (495.54,79.28) and (494.02,80.8) .. (492.14,80.8) .. controls (490.27,80.8) and (488.74,79.28) .. (488.74,77.4) -- cycle ;
\draw  [color=black  ,draw opacity=0 ][fill=black  ,fill opacity=1 ] (530.2,77.69) .. controls (530.2,75.81) and (531.72,74.29) .. (533.6,74.29) .. controls (535.47,74.29) and (537,75.81) .. (537,77.69) .. controls (537,79.57) and (535.47,81.09) .. (533.6,81.09) .. controls (531.72,81.09) and (530.2,79.57) .. (530.2,77.69) -- cycle ;

\draw (113.37,58.09) node [anchor=north west][inner sep=0.75pt]  [font=\tiny]  {$2$};
\draw (113.41,91.82) node [anchor=north west][inner sep=0.75pt]  [font=\tiny]  {$3$};
\draw (142,73.4) node [anchor=north west][inner sep=0.75pt]    { };
\draw (164.37,58.09) node [anchor=north west][inner sep=0.75pt]  [font=\tiny]  {$3$};
\draw (164.41,91.82) node [anchor=north west][inner sep=0.75pt]  [font=\tiny]  {$1$};
\draw (193,72.4) node [anchor=north west][inner sep=0.75pt]    { };
\draw (213.37,58.09) node [anchor=north west][inner sep=0.75pt]  [font=\tiny]  {$2$};
\draw (213.41,91.82) node [anchor=north west][inner sep=0.75pt]  [font=\tiny]  {$1$};
\draw (242,71.4) node [anchor=north west][inner sep=0.75pt]    { };
\draw (278.37,45.09) node [anchor=north west][inner sep=0.75pt]  [font=\tiny]  {$2$};
\draw (278.41,66.82) node [anchor=north west][inner sep=0.75pt]  [font=\tiny]  {$3$};
\draw (279.41,84.82) node [anchor=north west][inner sep=0.75pt]  [font=\tiny]  {$1$};
\draw (127,46.4) node [anchor=north west][inner sep=0.75pt]  [font=\tiny]  {$\lor $};
\draw (179,47.4) node [anchor=north west][inner sep=0.75pt]  [font=\tiny]  {$\lor $};
\draw (228,47.4) node [anchor=north west][inner sep=0.75pt]  [font=\tiny]  {$\lor $};
\draw (293,47.4) node [anchor=north west][inner sep=0.75pt]  [font=\tiny]  {$\lor $};
\draw (319,72.4) node [anchor=north west][inner sep=0.75pt]    { };
\draw (299,89.4) node [anchor=north west][inner sep=0.75pt]  [font=\scriptsize]  {\tiny$(1)$};
\draw (377,89.4) node [anchor=north west][inner sep=0.75pt]  [font=\scriptsize]  {\tiny$(2)$};
\draw (395,72.4) node [anchor=north west][inner sep=0.75pt]    { };
\draw (453,89.4) node [anchor=north west][inner sep=0.75pt]  [font=\scriptsize]  {\tiny$(3)$};
\draw (473,73.4) node [anchor=north west][inner sep=0.75pt]    { };
\draw (531,89.4) node [anchor=north west][inner sep=0.75pt]  [font=\scriptsize]  {\tiny$(4)$};
\draw (-25,66.4) node [anchor=north west][inner sep=0.75pt]    {$\{\widecheck{\varphi}_{j}^{\text{\tiny~(2-loop)}}\}_{j=1}^{7} =$};
\draw (353.37,45.09) node [anchor=north west][inner sep=0.75pt]  [font=\tiny]  {$2$};
\draw (353.41,66.82) node [anchor=north west][inner sep=0.75pt]  [font=\tiny]  {$3$};
\draw (354.41,84.82) node [anchor=north west][inner sep=0.75pt]  [font=\tiny]  {$1$};
\draw (368,47.4) node [anchor=north west][inner sep=0.75pt]  [font=\tiny]  {$\lor $};
\draw (430.37,44.5) node [anchor=north west][inner sep=0.75pt]  [font=\tiny]  {$2$};
\draw (430.41,66.82) node [anchor=north west][inner sep=0.75pt]  [font=\tiny]  {$3$};
\draw (431.41,84.82) node [anchor=north west][inner sep=0.75pt]  [font=\tiny]  {$1$};
\draw (445,46.4) node [anchor=north west][inner sep=0.75pt] [font=\tiny]   {$\lor $};
\draw (510.37,43.09) node [anchor=north west][inner sep=0.75pt]  [font=\tiny]  {$2$};
\draw (510.41,64.82) node [anchor=north west][inner sep=0.75pt]  [font=\tiny]  {$3$};
\draw (511.41,84.82) node [anchor=north west][inner sep=0.75pt]  [font=\tiny]  {$1$};
\draw (525,45.4) node [anchor=north west][inner sep=0.75pt]  [font=\tiny]  {$\lor $};
\end{tikzpicture}

%% file: tikz/lblBasis.tex
\tikzset{every picture/.style={line width=0.75pt}} 
\begin{tikzpicture}[x=0.75pt,y=0.75pt,yscale=-0.85,xscale=0.85]
\tikzset{
    orangerect/.style={
        draw=orange,
        fill=orange,
        opacity=0.5,
        rounded corners
    }
}
\begin{scope}[xshift=0,yshift=0]
\draw  [draw opacity=0][fill=violet  ,fill opacity=0.15 ] (285.69,20.84) .. controls (289.37,20.84) and (292.35,23.82) .. (292.35,27.5) -- (292.35,55.6) .. controls (292.35,59.28) and (289.37,62.26) .. (285.69,62.26) -- (265.7,62.26) .. controls (262.02,62.26) and (259.03,59.28) .. (259.03,55.6) -- (259.03,27.5) .. controls (259.03,23.82) and (262.02,20.84) .. (265.7,20.84) -- cycle;
\end{scope}
\draw  [draw opacity=0][fill=ForestGreen  ,fill opacity=0.15 ] (387.92,20.56) .. controls (392.5,20.56) and (396.2,24.27) .. (396.2,28.85) -- (396.2,53.7) .. controls (396.2,58.28) and (392.5,61.98) .. (387.92,61.98) -- (348.28,61.98) .. controls (343.71,61.98) and (340,58.28) .. (340,53.7) -- (340,28.85) .. controls (340,24.27) and (343.71,20.56) .. (348.28,20.56) -- cycle ;
\draw  [draw opacity=0][fill=Orange  ,fill opacity=0.15 ] (327.11,20.84) .. controls (330.79,20.84) and (333.77,23.82) .. (333.77,27.5) -- (333.77,55.6) .. controls (333.77,59.28) and (330.79,62.26) .. (327.11,62.26) -- (307.12,62.26) .. controls (303.44,62.26) and (300.45,59.28) .. (300.45,55.6) -- (300.45,27.5) .. controls (300.45,23.82) and (303.44,20.84) .. (307.12,20.84) -- cycle ;
\draw    (320.58,101.06) -- (327.96,101.08) ;
\draw  [draw opacity=0] (287.11,100.39) .. controls (287.11,100.39) and (287.11,100.39) .. (287.11,100.39) .. controls (287.11,100.39) and (287.11,100.39) .. (287.11,100.39) .. controls (287.11,90.41) and (294.68,82.31) .. (304.01,82.31) .. controls (313.34,82.31) and (320.9,90.41) .. (320.9,100.39) .. controls (320.9,100.49) and (320.9,100.59) .. (320.9,100.69) -- (304.01,100.39) -- cycle ; \draw  [color=Maroon  ,draw opacity=1 ] (287.11,100.39) .. controls (287.11,100.39) and (287.11,100.39) .. (287.11,100.39) .. controls (287.11,100.39) and (287.11,100.39) .. (287.11,100.39) .. controls (287.11,90.41) and (294.68,82.31) .. (304.01,82.31) .. controls (313.34,82.31) and (320.9,90.41) .. (320.9,100.39) .. controls (320.9,100.49) and (320.9,100.59) .. (320.9,100.69) ;  
\draw [color=black  ,draw opacity=1 ]   (384.63,45.35) -- (393.32,53.7);
\draw [color=black  ,draw opacity=1 ]   (341.09,54.61) -- (350.86,45.59) ;
\draw [color=black  ,draw opacity=1 ]   (304.06,54.95) -- (315.48,50.43) node[left,below,xshift=-14pt,yshift=2pt]{\tiny$,$};
\draw [color=black  ,draw opacity=1 ]   (315.48,50.43) -- (327.47,54.95) node[right]{\tiny$,$};
\begin{scope}[xshift=0,yshift=0]
\draw  [fill=black  ,fill opacity=1 ] (257.93,64.28) .. controls (255.66,64.25) and (254,54.34) .. (254.21,42.15) .. controls (254.42,29.97) and (256.43,20.12) .. (258.7,20.16) .. controls (256.64,21.46) and (254.91,30.8) .. (254.71,42.16) .. controls (254.51,53.52) and (255.92,62.91) .. (257.93,64.28) -- cycle ;
\end{scope}
\draw  [color=RoyalBlue  ,draw opacity=1 ]  (309.09,43.74) .. controls (305.38,40.07) and (305.45,34.18) .. (309.24,30.59) .. controls (313.03,27) and (319.11,27.06) .. (322.82,30.74) .. controls (326.52,34.41) and (326.46,40.3) .. (322.67,43.89) .. controls (322.67,43.89) and (322.67,43.89) .. (322.67,43.89) .. controls (318.09,48.22) and (315.8,50.39) .. (315.8,50.39) .. controls (315.8,50.39) and (313.57,48.18) .. (309.09,43.74) -- cycle ;
\draw  [draw opacity=0] (350.86,45.59) .. controls (350.86,45.59) and (350.86,45.59) .. (350.86,45.59) .. controls (350.86,35.61) and (358.42,27.52) .. (367.75,27.52) .. controls (377,27.52) and (384.51,35.48) .. (384.63,45.35) -- (367.75,45.59) -- cycle ; \draw  [color=RoyalBlue  ,draw opacity=1 ] (350.86,45.59) .. controls (350.86,45.59) and (350.86,45.59) .. (350.86,45.59) .. controls (350.86,35.61) and (358.42,27.52) .. (367.75,27.52) .. controls (377,27.52) and (384.51,35.48) .. (384.63,45.35) ;  
\draw  [draw opacity=0] (350.86,45.59) .. controls (350.86,54.59) and (358.42,61.89) .. (367.75,61.89) .. controls (376.99,61.89) and (384.5,54.72) .. (384.63,45.83) -- (367.75,45.59) -- cycle ; \draw  [color=RoyalBlue  ,draw opacity=1 ] (350.86,45.59) .. controls (350.86,54.59) and (358.42,61.89) .. (367.75,61.89) .. controls (376.99,61.89) and (384.5,54.72) .. (384.63,45.83) ;  
\draw    (342.18,37.24) -- (350.86,45.59) ;
\draw  [color=black  ,draw opacity=0 ][fill=black  ,fill opacity=1 ] (312.71,50.43) .. controls (312.71,48.95) and (313.95,47.75) .. (315.48,47.75) .. controls (317.01,47.75) and (318.25,48.95) .. (318.25,50.43) .. controls (318.25,51.92) and (317.01,53.12) .. (315.48,53.12) .. controls (313.95,53.12) and (312.71,51.92) .. (312.71,50.43) -- cycle ;
\draw  [color=black  ,draw opacity=0 ][fill=black  ,fill opacity=1 ] (348.09,45.59) .. controls (348.09,44.11) and (349.33,42.9) .. (350.86,42.9) .. controls (352.39,42.9) and (353.63,44.11) .. (353.63,45.59) .. controls (353.63,47.07) and (352.39,48.27) .. (350.86,48.27) .. controls (349.33,48.27) and (348.09,47.07) .. (348.09,45.59) -- cycle ;
\draw  [color=black  ,draw opacity=0 ][fill=black  ,fill opacity=1 ] (381.86,45.36) .. controls (381.86,43.87) and (383.1,42.67) .. (384.63,42.67) .. controls (386.16,42.67) and (387.4,43.87) .. (387.4,45.36) .. controls (387.4,46.84) and (386.16,48.04) .. (384.63,48.04) .. controls (383.1,48.04) and (381.86,46.84) .. (381.86,45.36) -- cycle ;
\draw    (98.88,108.26) -- (128.63,108.73) ;
\draw  [color=Maroon  ,draw opacity=1 ] (107.1,101.69) .. controls (103.42,97.94) and (103.49,91.91) .. (107.25,88.24) .. controls (111,84.56) and (117.03,84.63) .. (120.7,88.38) .. controls (124.38,92.14) and (124.31,98.17) .. (120.56,101.84) .. controls (116.02,106.28) and (113.75,108.5) .. (113.75,108.49) .. controls (113.75,108.5) and (111.54,106.23) .. (107.1,101.69) -- cycle ;
\begin{scope}[xshift=-2,yshift=-2]
\draw  [fill=black  ,fill opacity=1 ] (100.37,121.49) .. controls (98.15,121.49) and (96.35,111.47) .. (96.35,99.12) .. controls (96.35,86.76) and (98.15,76.74) .. (100.37,76.74) .. controls (98.38,78.1) and (96.83,87.6) .. (96.83,99.12) .. controls (96.83,110.63) and (98.38,120.13) .. (100.37,121.49) -- cycle ;
\end{scope}
\draw    (159.5,108.44) -- (189.9,108.82) ;
\draw  [color=Maroon  ,draw opacity=1 ] (168.37,101.79) .. controls (168.37,101.79) and (168.37,101.79) .. (168.37,101.79) .. controls (164.69,98.03) and (164.76,92) .. (168.52,88.33) .. controls (172.27,84.65) and (178.3,84.72) .. (181.97,88.48) .. controls (185.65,92.23) and (185.58,98.26) .. (181.82,101.93) .. controls (177.29,106.36) and (175.02,108.58) .. (175.02,108.59) .. controls (175.02,108.58) and (172.81,106.32) .. (168.37,101.79) -- cycle ;
\draw    (219.89,107.46) -- (249.65,107.92) ;
\draw  [color=Maroon  ,draw opacity=1 ] (228.11,100.89) .. controls (228.11,100.89) and (228.11,100.89) .. (228.11,100.89) .. controls (224.44,97.13) and (224.5,91.1) .. (228.26,87.43) .. controls (232.02,83.75) and (238.04,83.82) .. (241.72,87.58) .. controls (245.39,91.33) and (245.33,97.36) .. (241.57,101.03) .. controls (237.03,105.47) and (234.76,107.68) .. (234.77,107.69) .. controls (234.76,107.68) and (232.55,105.42) .. (228.11,100.89) -- cycle ;
\draw  [draw opacity=0] (287.11,100.82) .. controls (287.11,100.82) and (287.11,100.82) .. (287.11,100.82) .. controls (287.11,110.03) and (294.6,117.5) .. (303.85,117.5) .. controls (313.01,117.5) and (320.45,110.16) .. (320.58,101.06) -- (303.85,100.82) -- cycle ; \draw  [color=Maroon  ,draw opacity=1 ] (287.11,100.82) .. controls (287.11,100.82) and (287.11,100.82) .. (287.11,100.82) .. controls (287.11,110.03) and (294.6,117.5) .. (303.85,117.5) .. controls (313.01,117.5) and (320.45,110.16) .. (320.58,101.06) ;  
\draw    (279.73,100.8) -- (287.11,100.82) ;
\draw  [color=black  ,draw opacity=0 ][fill=RoyalBlue  ,fill opacity=1 ] (111.01,108.49) .. controls (111.01,106.98) and (112.24,105.75) .. (113.75,105.75) .. controls (115.27,105.75) and (116.5,106.98) .. (116.5,108.49) .. controls (116.5,110.01) and (115.27,111.24) .. (113.75,111.24) .. controls (112.24,111.24) and (111.01,110.01) .. (111.01,108.49) -- cycle ;
\draw  [color=black  ,draw opacity=0 ][fill=RoyalBlue  ,fill opacity=1 ] (171.95,108.63) .. controls (171.95,107.11) and (173.18,105.88) .. (174.7,105.88) .. controls (176.21,105.88) and (177.44,107.11) .. (177.44,108.63) .. controls (177.44,110.14) and (176.21,111.37) .. (174.7,111.37) .. controls (173.18,111.37) and (171.95,110.14) .. (171.95,108.63) -- cycle ;
\draw  [color=black  ,draw opacity=0 ][fill=RoyalBlue  ,fill opacity=1 ] (232.02,107.69) .. controls (232.02,106.17) and (233.25,104.94) .. (234.77,104.94) .. controls (236.28,104.94) and (237.51,106.17) .. (237.51,107.69) .. controls (237.51,109.2) and (236.28,110.43) .. (234.77,110.43) .. controls (233.25,110.43) and (232.02,109.2) .. (232.02,107.69) -- cycle ;
\draw  [color=black  ,draw opacity=0 ][fill=RoyalBlue  ,fill opacity=1 ] (284.37,100.82) .. controls (284.37,99.31) and (285.59,98.08) .. (287.11,98.08) .. controls (288.63,98.08) and (289.86,99.31) .. (289.86,100.82) .. controls (289.86,102.34) and (288.63,103.57) .. (287.11,103.57) .. controls (285.59,103.57) and (284.37,102.34) .. (284.37,100.82) -- cycle ;
\draw  [color=black  ,draw opacity=0 ][fill=RoyalBlue  ,fill opacity=1 ] (317.83,100.59) .. controls (317.83,99.07) and (319.06,97.84) .. (320.58,97.84) .. controls (322.09,97.84) and (323.32,99.07) .. (323.32,100.59) .. controls (323.32,102.1) and (322.09,103.33) .. (320.58,103.33) .. controls (319.06,103.33) and (317.83,102.1) .. (317.83,100.59) -- cycle ;
\draw    (384.63,45.36) -- (394.4,36.34) ;
\begin{scope}[xshift=0]
\draw  [fill=black  ,fill opacity=1 ] (397.39,20.12) .. controls (399.65,20.12) and (401.49,30) .. (401.49,42.19) .. controls (401.49,54.37) and (399.65,64.25) .. (397.39,64.25) .. controls (399.42,62.91) and (400.99,53.54) .. (400.99,42.19) .. controls (400.99,30.83) and (399.42,21.46) .. (397.39,20.12) -- cycle ;
\end{scope}
\draw    (304.06,46.84) -- (315.8,50.39) ;
\draw    (315.8,50.39) -- (327.47,46.84) ;
\draw [color=black  ,draw opacity=1 ]   (262.63,55.6) -- (274.06,51.08) ;
\draw [color=black  ,draw opacity=1 ]   (274.06,51.08) -- (286.05,55.6) ;
\draw  [color=RoyalBlue  ,draw opacity=1 ] (267.67,44.39) .. controls (263.96,40.72) and (264.03,34.83) .. (267.82,31.24) .. controls (271.61,27.65) and (277.69,27.71) .. (281.39,31.38) .. controls (285.1,35.06) and (285.04,40.94) .. (281.24,44.54) .. controls (281.24,44.54) and (281.24,44.54) .. (281.24,44.54) .. controls (276.68,48.87) and (274.38,51.05) .. (274.38,51.04) .. controls (274.38,51.05) and (272.14,48.83) .. (267.67,44.39) -- cycle ;
\draw  [color=black  ,draw opacity=0 ][fill=black  ,fill opacity=1 ] (271.29,51.08) .. controls (271.29,49.6) and (272.53,48.4) .. (274.06,48.4) .. controls (275.59,48.4) and (276.83,49.6) .. (276.83,51.08) .. controls (276.83,52.56) and (275.59,53.77) .. (274.06,53.77) .. controls (272.53,53.77) and (271.29,52.56) .. (271.29,51.08) -- cycle ;
\draw    (262.63,47.49) -- (274.38,51.04) ;
\draw    (274.38,51.04) -- (286.05,47.49) ;
\draw    (397.28,101.06) -- (404.66,101.08) ;
\draw  [draw opacity=0] (363.81,100.39) .. controls (363.81,100.39) and (363.81,100.39) .. (363.81,100.39) .. controls (363.81,100.39) and (363.81,100.39) .. (363.81,100.39) .. controls (363.81,90.41) and (371.38,82.31) .. (380.71,82.31) .. controls (390.04,82.31) and (397.6,90.41) .. (397.6,100.39) .. controls (397.6,100.49) and (397.6,100.59) .. (397.6,100.69) -- (380.71,100.39) -- cycle ; \draw  [color=Maroon  ,draw opacity=1 ] (363.81,100.39) .. controls (363.81,100.39) and (363.81,100.39) .. (363.81,100.39) .. controls (363.81,100.39) and (363.81,100.39) .. (363.81,100.39) .. controls (363.81,90.41) and (371.38,82.31) .. (380.71,82.31) .. controls (390.04,82.31) and (397.6,90.41) .. (397.6,100.39) .. controls (397.6,100.49) and (397.6,100.59) .. (397.6,100.69) ;  
\draw  [draw opacity=0] (363.81,100.82) .. controls (363.81,100.82) and (363.81,100.82) .. (363.81,100.82) .. controls (363.81,110.03) and (371.3,117.5) .. (380.55,117.5) .. controls (389.71,117.5) and (397.15,110.16) .. (397.28,101.06) -- (380.55,100.82) -- cycle ; \draw  [color=Maroon  ,draw opacity=1 ] (363.81,100.82) .. controls (363.81,100.82) and (363.81,100.82) .. (363.81,100.82) .. controls (363.81,110.03) and (371.3,117.5) .. (380.55,117.5) .. controls (389.71,117.5) and (397.15,110.16) .. (397.28,101.06) ;  
\draw    (356.43,100.8) -- (363.81,100.82) ;
\draw  [color=black  ,draw opacity=0 ][fill=RoyalBlue  ,fill opacity=1 ] (361.07,100.82) .. controls (361.07,99.31) and (362.3,98.08) .. (363.81,98.08) .. controls (365.33,98.08) and (366.56,99.31) .. (366.56,100.82) .. controls (366.56,102.34) and (365.33,103.57) .. (363.81,103.57) .. controls (362.3,103.57) and (361.07,102.34) .. (361.07,100.82) -- cycle ;
\draw  [color=black  ,draw opacity=0 ][fill=RoyalBlue  ,fill opacity=1 ] (394.53,100.59) .. controls (394.53,99.07) and (395.76,97.84) .. (397.28,97.84) .. controls (398.8,97.84) and (400.02,99.07) .. (400.02,100.59) .. controls (400.02,102.1) and (398.8,103.33) .. (397.28,103.33) .. controls (395.76,103.33) and (394.53,102.1) .. (394.53,100.59) -- cycle ;
\draw    (473.17,101.06) -- (480.56,101.08) ;
\draw  [draw opacity=0] (439.71,100.39) .. controls (439.71,100.39) and (439.71,100.39) .. (439.71,100.39) .. controls (439.71,90.41) and (447.27,82.31) .. (456.6,82.31) .. controls (465.93,82.31) and (473.5,90.41) .. (473.5,100.39) .. controls (473.5,100.49) and (473.5,100.59) .. (473.5,100.69) -- (456.6,100.39) -- cycle ; \draw  [color=Maroon  ,draw opacity=1 ] (439.71,100.39) .. controls (439.71,100.39) and (439.71,100.39) .. (439.71,100.39) .. controls (439.71,90.41) and (447.27,82.31) .. (456.6,82.31) .. controls (465.93,82.31) and (473.5,90.41) .. (473.5,100.39) .. controls (473.5,100.49) and (473.5,100.59) .. (473.5,100.69) ;  
\draw  [draw opacity=0] (439.71,100.82) .. controls (439.71,100.82) and (439.71,100.82) .. (439.71,100.82) .. controls (439.71,110.03) and (447.2,117.5) .. (456.44,117.5) .. controls (465.6,117.5) and (473.05,110.16) .. (473.17,101.06) -- (456.44,100.82) -- cycle ; \draw  [color=Maroon ,draw opacity=1 ] (439.71,100.82) .. controls (439.71,100.82) and (439.71,100.82) .. (439.71,100.82) .. controls (439.71,110.03) and (447.2,117.5) .. (456.44,117.5) .. controls (465.6,117.5) and (473.05,110.16) .. (473.17,101.06) ;  
\draw    (432.32,100.8) -- (439.71,100.82) ;
\draw  [color=black  ,draw opacity=0 ][fill=RoyalBlue  ,fill opacity=1 ] (436.96,100.82) .. controls (436.96,99.31) and (438.19,98.08) .. (439.71,98.08) .. controls (441.22,98.08) and (442.45,99.31) .. (442.45,100.82) .. controls (442.45,102.34) and (441.22,103.57) .. (439.71,103.57) .. controls (438.19,103.57) and (436.96,102.34) .. (436.96,100.82) -- cycle ;
\draw  [color=black  ,draw opacity=0 ][fill=RoyalBlue  ,fill opacity=1 ](470.43,100.59) .. controls (470.43,99.07) and (471.66,97.84) .. (473.17,97.84) .. controls (474.69,97.84) and (475.92,99.07) .. (475.92,100.59) .. controls (475.92,102.1) and (474.69,103.33) .. (473.17,103.33) .. controls (471.66,103.33) and (470.43,102.1) .. (470.43,100.59) -- cycle ;
\draw    (549.71,101.06) -- (557.1,101.08) ;
\draw  [draw opacity=0] (516.25,100.39) .. controls (516.25,100.39) and (516.25,100.39) .. (516.25,100.39) .. controls (516.25,100.39) and (516.25,100.39) .. (516.25,100.39) .. controls (516.25,90.41) and (523.81,82.31) .. (533.14,82.31) .. controls (542.47,82.31) and (550.04,90.41) .. (550.04,100.39) .. controls (550.04,100.49) and (550.04,100.59) .. (550.04,100.69) -- (533.14,100.39) -- cycle ; \draw  [color=Maroon  ,draw opacity=1 ] (516.25,100.39) .. controls (516.25,100.39) and (516.25,100.39) .. (516.25,100.39) .. controls (516.25,100.39) and (516.25,100.39) .. (516.25,100.39) .. controls (516.25,90.41) and (523.81,82.31) .. (533.14,82.31) .. controls (542.47,82.31) and (550.04,90.41) .. (550.04,100.39) .. controls (550.04,100.49) and (550.04,100.59) .. (550.04,100.69) ;  
\draw  [draw opacity=0] (516.25,100.82) .. controls (516.25,110.03) and (523.74,117.5) .. (532.98,117.5) .. controls (542.14,117.5) and (549.59,110.16) .. (549.71,101.06) -- (532.98,100.82) -- cycle ; \draw  [color=Maroon ,draw opacity=1 ] (516.25,100.82) .. controls (516.25,110.03) and (523.74,117.5) .. (532.98,117.5) .. controls (542.14,117.5) and (549.59,110.16) .. (549.71,101.06) ;  
\draw    (508.86,100.8) -- (516.25,100.82) ;
\draw  [color=black  ,draw opacity=0 ][fill=RoyalBlue  ,fill opacity=1 ] (513.5,100.82) .. controls (513.5,99.31) and (514.73,98.08) .. (516.25,98.08) .. controls (517.76,98.08) and (518.99,99.31) .. (518.99,100.82) .. controls (518.99,102.34) and (517.76,103.57) .. (516.25,103.57) .. controls (514.73,103.57) and (513.5,102.34) .. (513.5,100.82) -- cycle ;
\draw  [color=black  ,draw opacity=0 ][fill=RoyalBlue  ,fill opacity=1 ] (546.97,100.59) .. controls (546.97,99.07) and (548.2,97.84) .. (549.71,97.84) .. controls (551.23,97.84) and (552.46,99.07) .. (552.46,100.59) .. controls (552.46,102.1) and (551.23,103.33) .. (549.71,103.33) .. controls (548.2,103.33) and (546.97,102.1) .. (546.97,100.59) -- cycle ;
\begin{scope}[xshift=2,yshift=-2]
\draw  [fill=black  ,fill opacity=1 ] (585.3,74.94) .. controls (587.53,74.94) and (589.33,84.96) .. (589.33,97.32) .. controls (589.33,109.67) and (587.53,119.69) .. (585.3,119.69) .. controls (587.3,118.33) and (588.84,108.83) .. (588.84,97.32) .. controls (588.84,85.8) and (587.3,76.3) .. (585.3,74.94) -- cycle ;
\end{scope}
\draw (292.34,40.95) node [anchor=north west][inner sep=0.75pt]    { };
\draw (309.66,32.73) node [anchor=north west][inner sep=0.75pt]  [font=\tiny]  {$3$};
\draw (333.12,41.64) node [anchor=north west][inner sep=0.75pt]    { };
\draw (364.5,29.83) node [anchor=north west][inner sep=0.75pt]  [font=\tiny]  {$2$};
\draw (364.65,50.14) node [anchor=north west][inner sep=0.75pt]  [font=\tiny]  {$3$};
\draw (318.24,19.78) node [anchor=north west][inner sep=0.75pt]  [font=\tiny]  {$\lor $};
\draw (375.78,20.06) node [anchor=north west][inner sep=0.75pt]  [font=\tiny]  {$\lor $};
\draw (148.41,25.84) node [anchor=north west][inner sep=0.75pt]    {$\big\{\textcolor{RoyalBlue}{\widecheck{\varphi}_{i}^{\text{\tiny~(1-loop)}}}\big\}_{i=1}^{3} =$};
\draw (107.61,89.67) node [anchor=north west][inner sep=0.75pt]  [font=\tiny]  {$1$};
\draw (151.72,99.26) node [anchor=north west][inner sep=0.75pt]    { };
\draw (168.88,90.57) node [anchor=north west][inner sep=0.75pt]  [font=\tiny]  {$1$};
\draw (213.08,100.88) node [anchor=north west][inner sep=0.75pt]    { };
\draw (228.62,90.48) node [anchor=north west][inner sep=0.75pt]  [font=\tiny]  {$1$};
\draw (273.91,100.97) node [anchor=north west][inner sep=0.75pt]    { };
\draw (301.38,83.63) node [anchor=north west][inner sep=0.75pt]  [font=\tiny]  {$1$};
\draw (301.41,104.64) node [anchor=north west][inner sep=0.75pt]  [font=\tiny]  {$3$};
\draw (115.72,76.52) node [anchor=north west][inner sep=0.75pt]  [font=\tiny]  {$\lor $};
\draw (177.8,76.52) node [anchor=north west][inner sep=0.75pt]  [font=\tiny]  {$\lor $};
\draw (237.54,76.52) node [anchor=north west][inner sep=0.75pt]  [font=\tiny]  {$\lor $};
\draw (313.62,76.52) node [anchor=north west][inner sep=0.75pt]  [font=\tiny]  {$\lor $};
\draw (316.94,108.01) node [anchor=north west][inner sep=0.75pt]  [font=\scriptsize]  {\tiny$(1)$};
\draw (-17.39,83.68) node [anchor=north west][inner sep=0.75pt]    {$\big\{\textcolor{Maroon}{\widecheck{\varphi} _{ij}^{\text{\tiny~(left-over)}}}\big\}_{j=1}^{7} =$};
\draw (268.97,34.04) node [anchor=north west][inner sep=0.75pt]  [font=\tiny]  {$2$};
\draw (276.82,20.43) node [anchor=north west][inner sep=0.75pt]  [font=\tiny]  {$\lor $};
\begin{scope}[xshift=2pt,yshift=0pt]
\draw[fill=violet, rounded corners, fill opacity=0.15, draw opacity=0] (128.96,70.67) rectangle ++(0.45cm,1.2cm);
\draw (128.96,73.67) node [anchor=north west, inner sep=0.75pt] [font=\tiny]  
{$\begin{pmatrix}1\\0\\0\end{pmatrix}$\,,};
\end{scope}
\begin{scope}[xshift=2pt,yshift=0pt]
\draw[fill=orange, rounded corners, fill opacity=0.15, draw opacity=0] (190.03,71.67) rectangle ++(0.45cm,1.2cm);
\draw (190.03,73.67) node [anchor=north west, inner sep=0.75pt] [font=\tiny]  
{$\begin{pmatrix}0\\1\\0\end{pmatrix}$\,,};
\end{scope}
\begin{scope}[xshift=2pt,yshift=0pt]
\draw[fill=ForestGreen, rounded corners, fill opacity=0.15, draw opacity=0] (250.68,71.67) rectangle ++(0.45cm,1.2cm);
\draw (250.68,73.67) node [anchor=north west, inner sep=0.75pt] [font=\tiny]  
{$\begin{pmatrix}0\\0\\1\end{pmatrix}$\,,};
\end{scope}
\begin{scope}[xshift=2pt,yshift=0pt]
\draw[fill=ForestGreen, rounded corners, fill opacity=0.15, draw opacity=0] (330.98,71.67) rectangle ++(0.45cm,1.2cm);
\draw (330.98,73.67) node [anchor=north west, inner sep=0.75pt] [font=\tiny]  
{$\begin{pmatrix}0\\0\\1\end{pmatrix}$\,,};
\end{scope}
\draw (353.31,100.07) node [anchor=north west][inner sep=0.75pt]    { };
\draw (378.08,83.63) node [anchor=north west][inner sep=0.75pt]  [font=\tiny]  {$1$};
\draw (378.11,104.74) node [anchor=north west][inner sep=0.75pt]  [font=\tiny]  {$3$};
\draw (389.32,76.52) node [anchor=north west][inner sep=0.75pt]  [font=\tiny]  {$\lor $};
\draw (393.64,108.01) node [anchor=north west][inner sep=0.75pt]  [font=\scriptsize]  {\tiny$(2)$};
\begin{scope}[xshift=2pt,yshift=0pt]
\draw[fill=ForestGreen, rounded corners, fill opacity=0.15, draw opacity=0] (406.88,71.67) rectangle ++(0.45cm,1.2cm);
\draw (406.88,73.67) node [anchor=north west, inner sep=0.75pt] [font=\tiny]  
{$\begin{pmatrix}0\\0\\1\end{pmatrix}$\,,};
\end{scope}
\draw (427.41,100.07) node [anchor=north west][inner sep=0.75pt]    { };
\draw (453.98,83.63) node [anchor=north west][inner sep=0.75pt]  [font=\tiny]  {$1$};
\draw (454,104.54) node [anchor=north west][inner sep=0.75pt]  [font=\tiny]  {$3$};
\draw (465.31,76.52) node [anchor=north west][inner sep=0.75pt]  [font=\tiny]  {$\lor $};
\draw (469.53,108.01) node [anchor=north west][inner sep=0.75pt]  [font=\scriptsize]  {\tiny$(3)$};
\begin{scope}[xshift=2pt,yshift=0pt]
\draw[fill=ForestGreen, rounded corners, fill opacity=0.15, draw opacity=0] (483.57,71.67) rectangle ++(0.45cm,1.2cm);
\draw (483.57,73.67) node [anchor=north west, inner sep=0.75pt] [font=\tiny]  
{$\begin{pmatrix}0\\0\\1\end{pmatrix}$\,,};
\end{scope}
\draw (507.61,100.07) node [anchor=north west][inner sep=0.75pt]    { };
\draw (530.52,83.63) node [anchor=north west][inner sep=0.75pt]  [font=\tiny]  {$1$};
\draw (530.55,104.54) node [anchor=north west][inner sep=0.75pt]  [font=\tiny]  {$3$};
\draw (541.86,76.52) node [anchor=north west][inner sep=0.75pt]  [font=\tiny]  {$\lor $};
\draw (546.08,108.01) node [anchor=north west][inner sep=0.75pt]  [font=\scriptsize]  {\tiny$(4)$};
\begin{scope}[xshift=2pt,yshift=0pt]
\draw[fill=ForestGreen, rounded corners, fill opacity=0.15, draw opacity=0] (560.11,71.67) rectangle ++(0.45cm,1.2cm);
\draw (560.11,73.67) node [anchor=north west, inner sep=0.75pt] [font=\tiny]  
{$\begin{pmatrix}0\\0\\1\end{pmatrix}$};
\end{scope}

\end{tikzpicture}

%% file: skeleton.bbl
\providecommand{\href}[2]{#2}\begingroup\raggedright\begin{thebibliography}{10}

\bibitem{Bourjaily:2022bwx}
J.~L. Bourjaily et~al. 3, 2022.
\newblock \href{http://arxiv.org/abs/2203.07088}{{\tt arXiv:2203.07088}}.

\bibitem{Morales:2022csr}
R.~Morales et~al. {\em Phys. Rev. Lett.} {\bf 131} (2023), no.~4 041601, [\href{http://arxiv.org/abs/2212.09762}{{\tt arXiv:2212.09762}}].

\bibitem{Pogel:2022vat}
S.~P\"ogel, X.~Wang, and S.~Weinzierl {\em JHEP} {\bf 04} (2023) 117, [\href{http://arxiv.org/abs/2212.08908}{{\tt arXiv:2212.08908}}].

\bibitem{Duhr:2022dxb}
C.~Duhr et~al. {\em JHEP} {\bf 02} (2023) 228, [\href{http://arxiv.org/abs/2212.09550}{{\tt arXiv:2212.09550}}].

\bibitem{Duhr:2022pch}
C.~Duhr et~al. {\em Phys. Rev. Lett.} {\bf 130} (2023), no.~4 041602, [\href{http://arxiv.org/abs/2209.05291}{{\tt arXiv:2209.05291}}].

\bibitem{Cao:2023tpx}
Q.~Cao et~al. {\em JHEP} {\bf 04} (2023) 072, [\href{http://arxiv.org/abs/2301.07834}{{\tt arXiv:2301.07834}}].

\bibitem{Marzucca:2023gto}
R.~Marzucca, A.~J. McLeod, B.~Page, S.~P\"ogel, and S.~Weinzierl \href{http://arxiv.org/abs/2307.11497}{{\tt arXiv:2307.11497}}.

\bibitem{Bern:2023ccb}
Z.~Bern et~al. \href{http://arxiv.org/abs/2305.08981}{{\tt arXiv:2305.08981}}.

\bibitem{Henn:2013pwa}
J.~M. Henn {\em Phys. Rev. Lett.} {\bf 110} (2013) 251601, [\href{http://arxiv.org/abs/1304.1806}{{\tt arXiv:1304.1806}}].

\bibitem{Gorges:2023zgv}
L.~G\"orges et~al. {\em JHEP} {\bf 07} (2023) 206, [\href{http://arxiv.org/abs/2305.14090}{{\tt arXiv:2305.14090}}].

\bibitem{Caron-Huot:2021xqj}
S.~Caron-Huot and A.~Pokraka {\em JHEP} {\bf 12} (2021) 045, [\href{http://arxiv.org/abs/2104.06898}{{\tt arXiv:2104.06898}}].

\bibitem{Caron-Huot:2021iev}
S.~Caron-Huot and A.~Pokraka {\em JHEP} {\bf 04} (2022) 078, [\href{http://arxiv.org/abs/2112.00055}{{\tt arXiv:2112.00055}}].

\bibitem{Giroux:2022wav}
M.~Giroux and A.~Pokraka {\em JHEP} {\bf 03} (2023) 155, [\href{http://arxiv.org/abs/2210.09898}{{\tt arXiv:2210.09898}}].

\bibitem{Weinzierl:2022eaz}
S.~Weinzierl \href{http://arxiv.org/abs/2201.03593}{{\tt arXiv:2201.03593}}.

\bibitem{Smirnov:2010hn}
A.~V. Smirnov and A.~V. Petukhov {\em Lett. Math. Phys.} {\bf 97} (2011) 37--44, [\href{http://arxiv.org/abs/1004.4199}{{\tt arXiv:1004.4199}}].

\bibitem{Frellesvig:2019uqt}
H.~Frellesvig et~al. {\em Phys. Rev. Lett.} {\bf 123} (2019), no.~20 201602, [\href{http://arxiv.org/abs/1907.02000}{{\tt arXiv:1907.02000}}].

\bibitem{giroux:2023}
M.~Giroux, A.~Pokraka, F.~Porkert, and Y.~Sohnle \emph{Work in progress}.

\bibitem{hjalteTalk}
H.~Frellesvig Talk at \textsc{RADCOR 2023}, available \href{https://indico.ph.ed.ac.uk/event/118/contributions/2332/attachments/1181/1678/Hjalte.pdf}{here}.

\bibitem{Bogner:2019lfa}
C.~Bogner et~al. {\em Nucl. Phys. B} {\bf 954} (2020) 114991, [\href{http://arxiv.org/abs/1907.01251}{{\tt arXiv:1907.01251}}].

\end{thebibliography}\endgroup
